\documentclass[manuscript]{aastex}

\shorttitle{Temporary Capture of Asteroids by a Planet}
\shortauthors{Higuchi \& Ida}

\begin{document}

\title{Temporary Capture of Asteroids by a Planet:
Dependence of Prograde/Retrograde Capture on Asteroids' Semimajor Axes}
\author{A. Higuchi}
\affil{Department of Earth and Planetary Sciences, Faculty of Science, 
  Tokyo Institute of Technology, Meguro, Tokyo 152-8551, Japan}
\and
\author{S. Ida}
\affil{Earth-Life Science Institute,
  Tokyo Institute of Technology, Meguro, Tokyo 152-8550, Japan}


\begin{abstract}
We have investigated the dependence of the prograde/retrograde temporary capture of 
 asteroids by a planet on their original heliocentric semimajor axes
 through analytical arguments and numerical orbital integrations
in order to discuss the origins of irregular satellites of giant planets.
We found that capture is mostly retrograde for the asteroids 
 near the planetary orbit and is prograde for those from further orbits.
An analytical investigation reveals the intrinsic dynamics of these dependences and
gives boundary semimajor axes for the change in prograde/retrograde capture.
The numerical calculations support the idea of deriving the analytical formulae
and confirm their dependence.
Our numerical results show that the capture probability 
  is much higher for bodies from the inner region than for outer ones.
These results imply that retrograde irregular satellites of Jupiter
  are most likely to be captured bodies from the nearby orbits of Jupiter
  that may have the same origin as Trojan asteroids,
  while prograde irregular satellites originate from far inner regions
  such as the main-belt asteroid region.
\end{abstract}

\keywords{planets and satellites: formation}

\section{Introduction}

Irregular satellites around giant planets are small satellites
 with elliptical and inclined orbits
 \citep[e.g.,][]{jh07, n08}.
They have relatively large (planetocentric) semimajor axes.
Because of their orbits, they are usually thought to be captured passing asteroids
rather than formed {\it in situ}.
In some cases, when the velocity of an asteroid
 relative to the planet is relatively low,
 it is temporarily trapped in the planetary Hill sphere.
The trapped body must eventually exit the Hill sphere.
But, if some energy loss (e.g., tidal dissipation, 
 drag force from a circumplanetary disk when it existed,
 or collisions with other solid bodies in the disk) 
 affects the asteroid's orbit,
 it can be permanently captured afterward. 
In fact, such temporary capture events have been observed.
Examples include 
(1) comet D/Shoemaker-Levy 9,
 which impacted Jupiter \citep{ww95},
(2) 2006RH$_{120}$, which spent some time
    within Earth's Hill radius and was observed during the capture
    \citep[e.g.,][]{kk09}, and
(3) some bodies, including 147/Kushida-Muramatsu,
    which were found to be temporarily captured by Jupiter
    through backward orbital integrations \citep{oiyaa08}.

  Many studies on the origins of irregular satellites
  have been published \citep[e.g.,][]{kd96, abwf03, cb04, n07, p10, sot11, n14}.
  However, they mainly used numerical orbital integrations 
  in the restricted three-body problem or more complex framework
  (e.g., with gas drag or perturbations from other objects),
  so it is not easy to understand general relationships between
  the original heliocentric orbits of the captured asteroids
  and their planetocentric orbits as satellites.

In this study, we approximate a circular
 three-body problem (Sun-planet-asteroid) into 
  a combination of two independent two-body problems (Sun-asteroid and planet-satellite),
  identifying the asteroid with the satellite,
  and derive the relation between the pre-capture heliocentric 
  orbit and the planetocentric orbit at the moment of capture.
We derive analytical formulae with simple assumptions 
  and show that the formulae 
  reproduce the results of orbital integrations very well.
 The analytical formulae reveal the intrinsic dynamics that regulates
  the relation between the heliocentric and planetocentric orbital elements.
 In particular, we show a clear dependence of prograde vs. retrograde
 capture of Jupiter's irregular satellites on 
  the heliocentric semimajor axes of the original asteroids.
 
 We describe the assumptions, basic formulation, derivation
    of the analytical formulae, and show orbital distributions
    generated by the formulae in Section 2.
    The methods and results of numerical calculations are presented in Section 3.
    In Section 4, we summarize the results and discuss the origin of Jupiter's 
    irregular satellites,
    referring to results of the photometric observations.

\section{Analytical Prediction for Temporary Capture}

We first derive analytical formulae for temporary capture.
Next, we generate the distribution of planetocentric orbits of the
 temporarily captured satellites using the formulae.

\subsection{Derivation of Formulae for Temporary Capture}

\subsubsection{Assumptions}
We use four conditions for temporary capture.
We split a circular three-body problem (Sun-planet-asteroid) into 
 two independent two-body problems (Sun-asteroid and planet-asteroid),
 identifying the asteroid with the satellite.
We use the relative velocity between the asteroid and the planet
in heliocentric orbits as a satellite velocity orbiting around the planet 
 (condition [1])
 at the $L_1$ or $L_2$ Lagrangian point of the planet,
 which is a Hill radius away from the planet on the $x-$axis
 (condition [2]).
The Hill radius is defined by $r_{\rm H}=a_{\rm p}(m_{\rm p}/3M_\odot)^{1/3}$,
 where $a_{\rm p}$ and $m_{\rm p}$ are the semimajor axis and mass of the 
 planet and $M_\odot$ is the solar mass.
Entering the zero-velocity surface that surrounds the planet
 via the $L_1$ or $L_2$ points provides the easiest access
 to planetocentric orbits
 in the restricted three-body problem \citep[e.g.,][]{md99}.

Additionally, we set two other conditions to make the derivation simpler:
 we assume that the body's position
 at the moment of transition from heliocentric motion to planetocentric motion, 
 (i.e., the $L_1$ or $L_2$ point) 
 is the aphelion or perihelion 
 of the heliocentric orbit (condition [3]).
Condition [3], which implies that the relative radial velocity is zero,
 leads to the condition that the body has 
 its apocenter or pericenter 
 on the planetocentric orbit at $L_1$ or $L_2$ (condition [4]).
An apocenter at $L_1$ or $L_2$ corresponds
 to a planetocentric orbit within $r_{\rm H}$ from the planet.
A temporary capture does not always start with such a tightly bound orbit,
 so we relax the condition to having either an apocenter \emph{or} pericenter
 at $L_1$ or $L_2$.
Condition [4] is not always a good approximation because the planet's 
 gravitational pull
 creates a velocity component radial to the planet.
However, condition [3] is usually an acceptable approximation, so that
 condition [4] is necessary for the identification between heliocentric and planetocentric motions.

\subsubsection{Conditions for Temporary Capture}

We consider a body and a planet in the rotating Cartesian coordinates
($x$, $y$, $z$) centered on the Sun.
The $x$-axis is chosen as parallel to the planet's position vector
 from the Sun and
the $x$-$y$ plane lies in the planet's orbital plane.

The heliocentric velocity of the body (an asteroid) at the heliocentric distance $r$ is
\begin{eqnarray}
  v&=&\sqrt{GM_\odot\left(\frac{2}{r}-\frac{1}{a}\right)},
  \label{eq:tc_v}
\end{eqnarray}
 where $G$ is the gravitational constant, $M_\odot$ is the solar mass, and 
 $a$ is the heliocentric semimajor axis of the body.
Using Equation (\ref{eq:tc_v}) and conditions [2] and [3],
we obtain the velocity at the orbit transition from heliocentric to planetocentric orbit as 
\begin{eqnarray}
    v = \sqrt{GM_\odot\left(\frac{2}{A_\mp a_{\rm p}}-\frac{1}{a}\right)}
    = v_{\rm p}\chi, 
    \label{eq:tc_v2}
    \\
    \chi =  \sqrt{\frac{2}{A_\mp}-\frac{a_{\rm p}}{a}},
    \label{eq:tc_chi}
    \\
    \left\{
    \begin{array}{l}
      A_- = 1-\hat{r}_{\rm H}\;\;{\rm at}\;\;L_1\\
      A_+ = 1+\hat{r}_{\rm H}\;\;{\rm at}\;\;L_2,
    \end{array}
    \right.
    \label{eq:tc_aa}
\end{eqnarray}
where $v_{\rm p}=\sqrt{GM_\odot/a_{\rm p}}$ and $\hat{r}_{\rm H}=r_{\rm H}/a_{\rm p}$.

From condition [3], the velocity vector is written as 
 $(v_x, v_y, v_z)=(0, v\cos i, v\sin i)$,
 where $i$ is the heliocentric orbital inclination of the body
 measured from the orbital plane of the planet.
The planet has a heliocentric velocity ${\bf v}_{\rm p}=(0,v_{\rm p}, 0)$.

The velocity of the body (a satellite) orbiting around the planet is written as
\begin{eqnarray}
  v_{\rm s}&=&\sqrt{Gm_p\left(\frac{2}{r_{\rm s}}-\frac{1}{a_{\rm s}}\right)},
  \label{eq:tc_vs}
\end{eqnarray}
 where $r_{\rm s}$ and $a_{\rm s}$ are the planetocentric distance
 and semimajor axis of the body.
From condition [4], 
\begin{eqnarray}
  r_{\rm s}=a_{\rm s}\frac{1-e_{\rm s}^2}{1+e_{\rm s}\cos f_{\rm s}}=r_{\rm H},
  \label{eq:tc_rs}
\end{eqnarray}
where $e_{\rm s}$ and $f_{\rm s}$ are the eccentricity (satellite eccentricity)
 and the true anomaly of the body
 in the planetocentric orbit.

Substituting Equation (\ref{eq:tc_rs}) into (\ref{eq:tc_vs}),
we obtain 
\begin{eqnarray}
  v_{\rm s}=v_{\rm p}\sqrt{\frac{m_{\rm p}}{M_\odot}\frac{\kappa}{\hat{r}_{\rm H}}}
  =\sqrt{3\kappa}v_{\rm p} \hat{r}_{\rm H},
  \label{eq:tc_vsat}
  \\
  \kappa = \frac{1+e_{\rm s}^2+2e_{\rm s}\cos f_{\rm s}}{1+e_{\rm s}\cos f_{\rm s}},
  \label{eq:tc_kappag}
\end{eqnarray}
If condition [4] is satisfied, 
Equation (\ref{eq:tc_kappag}) becomes a function only of $e_{\rm s}$,
\begin{eqnarray}
  \left\{
  \begin{array}{l}
    \kappa = 1-e_{\rm s}\;\;{\rm at\;planetocentric\;apocenter}\\
    \kappa = 1+e_{\rm s}\;\;{\rm at\;planetocentric\;pericenter},
  \end{array}
  \right.
  \label{eq:tc_kappa}
\end{eqnarray}
and
\begin{eqnarray}
  r_{\rm s}=a_{\rm s}\frac{1-e_{\rm s}^2}{1+e_{\rm s}\cos f_{\rm s}}=r_{\rm H},
  \label{eq:tc_rsg}
\end{eqnarray}
where $f_{\rm s}=0$ and $f_{\rm s}=180^\circ$ are substituted at apocenter and pericenter, 
 respectively.
We use $\kappa$ defined by Equation (\ref{eq:tc_kappa}) as a parameter.

\subsection{Heliocentric Orbital Elements for Temporary Capture -Planar Case}
\label{ss:tc_p} 

First, we consider the simplest planar case with $i = 0$
 to understand how the orbital character (prograde or retrograde)
 is determined by the original heliocentric semimajor axis of a temporarily captured body. 
We will discuss general cases with non-zero $i$ later. 
However, as we will show, for the body to be captured, 
 the heliocentric orbital inclination must be relatively small ($i\lesssim 10^\circ$)
 so that the discussion of the planar case can be generalized to understand
 all other cases. 
   
The capture would be prograde if $v_{\rm rel} = v_y - v_{\rm p} < 0$ at $L_1$
 or $v_{\rm rel}  > 0$ at $L_2$.
Otherwise, the capture would be retrograde. 
From Equation (\ref{eq:tc_v2}),
\begin{eqnarray}
  v_{\rm rel}  = v_{\rm p}
  \left(\chi-1\right).
  \label{eq:tc_x}
\end{eqnarray}
This equation implies that the capture is 
\begin{eqnarray}
 \left\{
   \begin{array}{lccl}
    {\displaystyle 
    \mbox{prograde if }
    \frac{a}{a_{\rm p}} < B_- 
    }
    &{\rm and}&
    {\displaystyle 
    \mbox{retrograde if }
    \frac{a}{a_{\rm p}} > B_-
    }
      & \mbox{[at } L_1], 
    \\
    {\displaystyle 
    \mbox{retrograde if }
    \frac{a}{a_{\rm p}} < B_+ }
    &{\rm and}&
   {\displaystyle 
    \mbox{retrograde if }
    \frac{a}{a_{\rm p}} > B_+ 
    }
    & \mbox{[at } L_2],
    \label{eq:tc_xx}
  \end{array}
\right.
\end{eqnarray}
 where 
\begin{eqnarray}
  B_-=\frac{1-\hat{r}_{\rm H}}{1+\hat{r}_{\rm H}}
  \;\;{\rm and}\;\;
  B_+=\frac{1+\hat{r}_{\rm H}}{1-\hat{r}_{\rm H}}.
  \label{eq:tc_bb}
\end{eqnarray}
We always find that
$B_-< 1 < B_+$ and the difference between $B_-$ and $B_+$ 
 is larger for bigger $m_{\rm p}$ (larger $\hat{r}_{\rm H}$).

If the relative velocity $| v_{\rm rel} |$
 exceeds the velocity limit for bounded orbits around the planet,
 the body cannot be temporarily captured. 
As Equation (\ref{eq:tc_x}) shows, $v_{\rm rel}$
 takes a large negative value when $a$ is too small 
 (i.e., $\chi\rightarrow 0$)
 and a large positive value when $a$ is too large 
 (i.e., $\chi\rightarrow \sqrt{2/(1\pm\hat{r}_{\rm H})}$).
Thus, temporary capture is possible for bodies with heliocentric
 semimajor axes in some range encompassing $a_{\rm p}$.
We will derive the range of $a$ in the following.
 
During the transition from a heliocentric orbit to a planetocentric orbit,
$v_{\rm rel}  = \pm v_{\rm s}$ for some value of $\kappa$.
For the body to have an orbit bounded to the planet,
$e_{\rm s} < 1$, that is, $0 \le \kappa < 2$.
Using Equations (\ref{eq:tc_vsat}) and (\ref{eq:tc_x}),
\begin{eqnarray}
  \chi -1
  = \mp (3\kappa)^{1/2} \hat{r}_{\rm H} 
  \label{eq:tc_11}
\end{eqnarray}
 where $-$ and $+$ in the rhs
 correspond to $a < a_{\rm p}$ and $a > a_{\rm p}$, respectively,
 and are independent of $\mp$ in $A_\mp$.
Using Equation (\ref{eq:tc_chi}),
 Equation (\ref{eq:tc_11}) is rewritten as
\begin{eqnarray}
  \frac{a}{a_{\rm p}}
  &=&\left\{
    \frac{2}{A_\mp}
    -\left[1 \mp 
      (3\kappa)^{\frac{1}{2}}\hat{r}_{\rm H}\right]^2
  \right\}^{-1}.
  \label{eq:tc_a_mnmx}
\end{eqnarray}

Thus, both the minimum and the maximum values of $a$ are given
 with $\kappa =2$ :
\begin{eqnarray}
    \frac{a_{\rm min \mp}}{a_{\rm p}} \;=\; C_\mp
    &=&\left\{
    \frac{2}{A_\mp}
    -\left[1 -
      \sqrt{6}\hat{r}_{\rm H}\right]^2
    \right\}^{-1},
    \label{eq:tc_xxx1}
    \\
    \frac{a_{\rm max \mp}}{a_{\rm p}} \;=\;D_\mp
    &=&\left\{
    \frac{2}{A_\mp}
    -\left[1 +
      \sqrt{6}\hat{r}_{\rm H}\right]^2
    \right\}^{-1},
  \label{eq:tc_xxx2}
\end{eqnarray}
 where $-$ and $+$ correspond to expressions with $A_- = 1-\hat{r}_{\rm H}$
 and those with $A_+ =1+\hat{r}_{\rm H}$, respectively.
Note that $C_- < C_+ < 1 < D_- < D_+$. 
Combining Equations (\ref{eq:tc_xx}), (\ref{eq:tc_xxx1}), and (\ref{eq:tc_xxx2}),
 we find that $L_1$ capture is
\begin{eqnarray}
 \left\{
  \begin{array}{lcc}
    &\mbox{prograde for}&
    {\displaystyle 
    C_-<\frac{a}{a_{\rm p}} < B_- ,}
    \\
    &\mbox{retrograde for}&
     {\displaystyle 
    B_-<\frac{a}{a_{\rm p}} < D_- ,}
  \end{array}
  \right.
\end{eqnarray}
while $L_2$ capture is
\begin{eqnarray}
\left\{
  \begin{array}{lcc}
    &\mbox{retrograde for}&
     {\displaystyle 
    C_+<\frac{a}{a_{\rm p}} < B_+ ,}
    \\
    &\mbox{prograde for}&
     {\displaystyle 
    B_+<\frac{a}{a_{\rm p}} < D_+ .}
  \end{array}
  \right.
\end{eqnarray}
In other words, the capture direction is summarized as
\begin{eqnarray}
  \begin{array}{lcc}
    (1)\; \mbox{only prograde} 
    &\mbox{ for }&
    C_- < a/a_{\rm p} < C_+, 
    \\
    (2)\; \mbox{prograde via $L_1$ and retrograde via $L_2$}   
    &\mbox{ for }&
    C+ < a/a_{\rm p} < B_- ,
    \\
    (3)\; \mbox{only retrograde}
    &\mbox{ for }&
    B_- < a/a_{\rm p} < B_+,
    \\
    (4)\; \mbox{prograde via $L_2$ and retrograde via $L_1$}   
    &\mbox{ for }&
    B_+ < a/a_{\rm p} < D_- ,
    \\
    (5)\; \mbox{only prograde}
    &\mbox{ for }&
    D_- < a/a_{\rm p} < D_+ , 
  \end{array}
  \label{eq:tc_12345}
\end{eqnarray}
where 
$B_\pm$ is defined by Equation (\ref{eq:tc_bb}) and 
$C_\mp$ and $D_\mp$ are defined by Equations (\ref{eq:tc_xxx1}) and (\ref{eq:tc_xxx2}),
 respectively. 
For example, for $m_{\rm p} = 9.55\times 10^{-4}M_\odot$ and $a_{\rm p}=5.2$ AU, 
$\hat{r}_{\rm H} = 0.068$ and $C_- a_{\rm p} = 3.6$ AU, $C_+ a_{\rm p} =4.4$ AU,
$B_- a_{\rm p} = 4.5$ AU, $B_+ a_{\rm p} = 6.0$ AU, 
$D_- a_{\rm p} = 6.6$ AU, and $D_+ a_{\rm p} =10.2$ AU.
Therefore, whether prograde or retrograde capture dominates
 clearly depends on the heliocentric semimajor axis of the captured body. 
The ratio of prograde capture to retrograde capture in
 the ranges of (2) and (4) is determined by the ratio of capture
 via $L_1$ to that via $L_2$, depending on the distributions of 
 heliocentric orbital elements of asteroids.

\subsection{Heliocentric Orbital Elements for Temporary Capture -General Case}

\subsubsection{Generalized Condition for Temporary Capture}
Now we extend the discussions in section \ref{ss:tc_p} to the general
 case by including inclined (asteroidal) orbits.
The velocity at the orbit transition from a heliocentric to planetocentric orbit 
 with  heliocentric orbital inclination $i$ 
 is $(v_x, v_y, v_z)=(0, v_{\rm p}\chi\cos i, v_{\rm p}\chi\sin i)$.
Then, the relative velocity of the body to the planet is written as 
\begin{eqnarray}
  v_{\rm rel}
  &=&\left(v_{{\rm rel}, y}^2 + v_{{\rm rel}, z}^2\right)^{\frac{1}{2}}
  \nonumber\\
  &=&\left[(v_y -v_{\rm p})^2 + v_z^2\right]^{\frac{1}{2}}
  \nonumber\\
  &=&v_{\rm p}\left[\chi^2 +1-2\chi\cos i \right]^{\frac{1}{2}}.
  \label{eq:tc_rvk}
\end{eqnarray}
With
$|v_{\rm rel}|= v_{\rm s}$,
 the equation of temporary capture can be written as follows:
\begin{eqnarray}
  \chi^2-2\chi\cos i+1 -3\kappa\hat{r}_{\rm H}^2=0.
  \label{eq:tc_hokaku}
\end{eqnarray}

Equation (\ref{eq:tc_hokaku}) 
 connects the incoming heliocentric orbital elements before the 
 capture ($a, i$) with the planetocentric orbital elements at the moment of capture
 ($a_{\rm s}$, $e_{\rm s}$, $i_{\rm s}$).
Equation (\ref{eq:tc_hokaku}) also gives the range of these orbital parameters
 that satisfy the conditions of temporary capture by substituting $0\le \kappa < 2$.
As already shown in Equation (\ref{eq:tc_kappa}), 
$\kappa$ is a parameter that shows the position along the 
 planetocentric orbit at the moment of capture.
The body captured at apocenter corresponds to $0<\kappa<1$,
 and to $1<\kappa<2$ when captured at pericenter.
The planetocentric orbit of a body captured at pericenter 
 has its apocenter at $\kappa r_{\rm H}$, 
 which is outside the Hill radius.
In general, temporary capture that results in a tightly bound 
 orbit within planet's Hill radius corresponds to 
 $0 \le \kappa < 1$ rather than $1 \le \kappa < 2$.

\paragraph{Range of the Semimajor Axis.}

The $a$ range for temporary capture is given by solving Equation (\ref{eq:tc_hokaku})
 and substituting Equation (\ref{eq:tc_chi}) into the solution.
We use an $a_{\rm min}$ and $a_{\rm max}$ of
\begin{eqnarray}
  \frac{a_{\rm min \mp}}{a_{\rm p}}
  &=&\left\{
    \frac{2}{A_\mp}
    -\left[\cos i -\sqrt{\cos^2 i -1
	+6\hat{r}_{\rm H}^2}\right]^2
  \right\}^{-1},
  \label{eq:tc_a_mn}
  \\
  \frac{a_{\rm max \mp}}{a_{\rm p}}
  &=&\left\{
    \frac{2}{A_\mp}
    -\left[\cos i + \sqrt{\cos^2 i -1
	+6\hat{r}_{\rm H}^2}\right]^2
  \right\}^{-1}.
  \label{eq:tc_a_mx}
\end{eqnarray}
These are generalized forms of Equations (\ref{eq:tc_xxx1}) and (\ref{eq:tc_xxx2}).
The values of $a_{\rm min}$ and $a_{\rm max}$ for $i=0$ for eight planets 
 are given in Table \ref{tb:tc_range}.
Equations (\ref{eq:tc_a_mn}) and (\ref{eq:tc_a_mx}) show that the
 $a-$ range for temporary capture becomes largest when $i=0$.
This is because $v_{\rm rel}$ is smaller for smaller $i$ and $a\sim a_{\rm p}$ 
 (Eq. (\ref{eq:tc_rvk})).
In other words, small $i$ is required for a body with $a$ far from the planet to be captured.

\begin{figure}[hbtp]
  \begin{center}
    \resizebox{13cm}{!}{\includegraphics{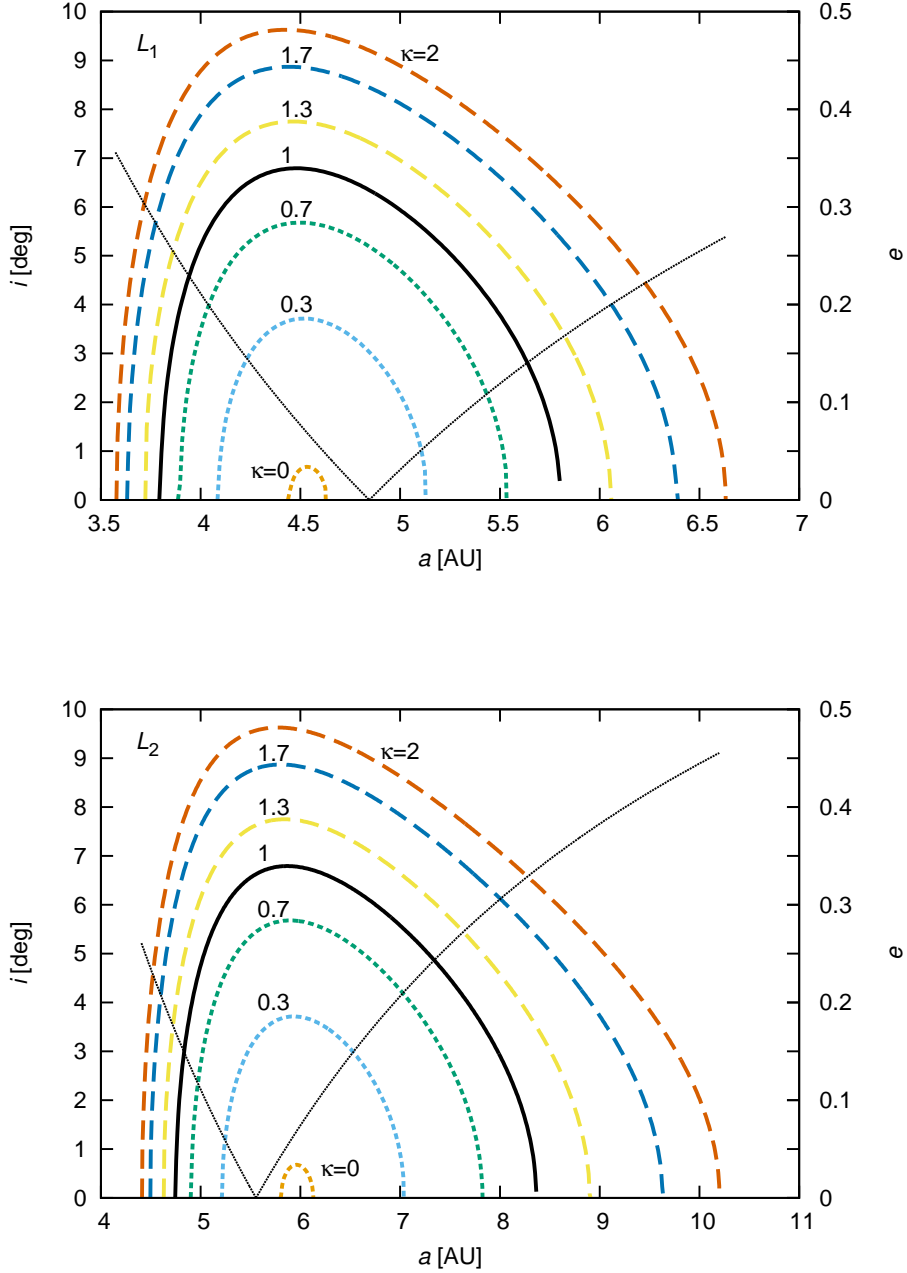}}
    \caption{Solutions to Equation (\ref{eq:tc_hokaku}) on the $a-i$ plane for various $\kappa$.
      The upper and lower panels are for $L_1$ and $L_2$ capture, respectively.
      The numbers labeled on the curves represent the values of $\kappa$.
      The eccentricity $e$ given for each $a$ by condition [3] is also plotted against
      the secondary (right) $y-$axis (thin dashed curve).
      The upper and lower panels are for $L_1$ and $L_2$ capture, respectively.}
    \label{fig:tc_a_i}
  \end{center}
\end{figure}

\paragraph{Range of Inclination.}
 Figure \ref{fig:tc_a_i} shows solutions to Equation (\ref{eq:tc_hokaku}) on
 the $a-i$ plane, for Jupiter with $a_{\rm p}=5.2$AU and $m_{\rm p}=9.55\times 10^{-4}M_\odot$
 for $0\le\kappa\le 2$.
For these planetary parameters, $L_1$ and $L_2$ are located
 at 4.84 and 5.56AU, respectively. 
The two feet of the curves in each panel touching the $x-$axis for $\kappa=2$ 
 correspond to $C_- a_{\rm p}$, $C_+ a_{\rm p}$, $D_- a_{\rm p}$, and $D_+ a_{\rm p}$.
As shown in the derivation of Equations (\ref{eq:tc_a_mn}) and (\ref{eq:tc_a_mx}),
 $C_- a_{\rm p}$ and $D_+ a_{\rm p}$ are $a_{\rm min}$ and $a_{\rm max}$ for $i=0$, respectively.

The region where $a$ has a real value in Equation (\ref{eq:tc_hokaku})
 is defined by the region enclosed with 
 the curve for $\kappa=2$ 
 and the $x$-axis in Figure \ref{fig:tc_a_i}.
Hereafter, we call this region as the TC region.

The maximum inclination in the TC region is given by
\begin{eqnarray}
  i_{\rm max} = 
  {\rm arccos}\left(\sqrt{1-6\hat{r}^2_{\rm H}}\right),
  \label{eq:tc_imax}
\end{eqnarray}
This is a function only of the mass of the planet.
For $m_{\rm p}=9.55\times 10^{-4}M_\odot$, 
$i_{\rm max}\simeq 9.6^\circ$.
The values of $i_{\rm max}$ for the eight planets are listed in Table \ref{tb:tc_range}.

\begin{figure}[hbtp]
  \begin{center}
    \resizebox{13cm}{!}{\includegraphics{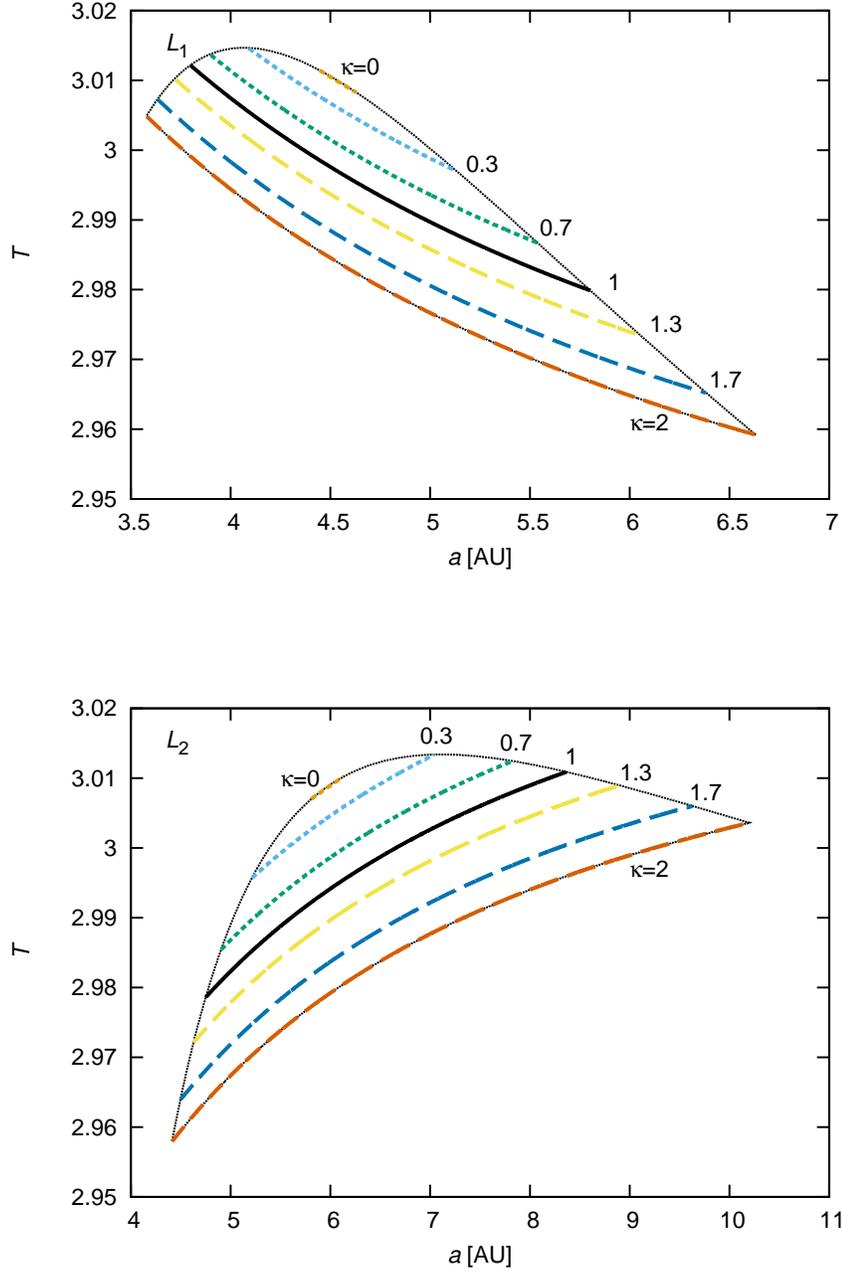}}
    \caption{Solutions to Equation (\ref{eq:tc_hokaku}) on the $a-T$ plane for various $\kappa$.
      The upper and lower panels are for $L_1$ and $L_2$ capture, respectively.
      The numbers labeled on the curves represent the values of $\kappa$.
      The black envelope shows the TC regions filled with curves for $0<\kappa<2$.
    }
    \label{fig:tc_a_T}
  \end{center}
\end{figure}

\paragraph{Range of the Tisserand Parameter.}
Figure \ref{fig:tc_a_T} shows the TC region on the $a-T$ plane where $T$ is the Tisserand parameter 
defined as
\begin{eqnarray}
  T=\frac{a_{\rm p}}{a} +2\sqrt{\frac{a}{a_{\rm p}}(1-e^2)}\cos i.
\end{eqnarray}
The Tissserand parameter shows the orbital relation between the body
  and the planet in the circular restricted three-body framework:
  bodies with $T>3$ never cross the planetary orbit.
So the existence of the TC regions for $T>3$ means that planets can capture
 bodies whose orbits are not potentially planetary orbit-crossing.
The minimum and maximum values of $T$ for the eight planets are numerically computed
 and given in Table \ref{tb:tc_range}.

\subsubsection{Satellite inclination}
\begin{figure}[hbtp]
  \begin{center}
    \resizebox{13cm}{!}{\includegraphics{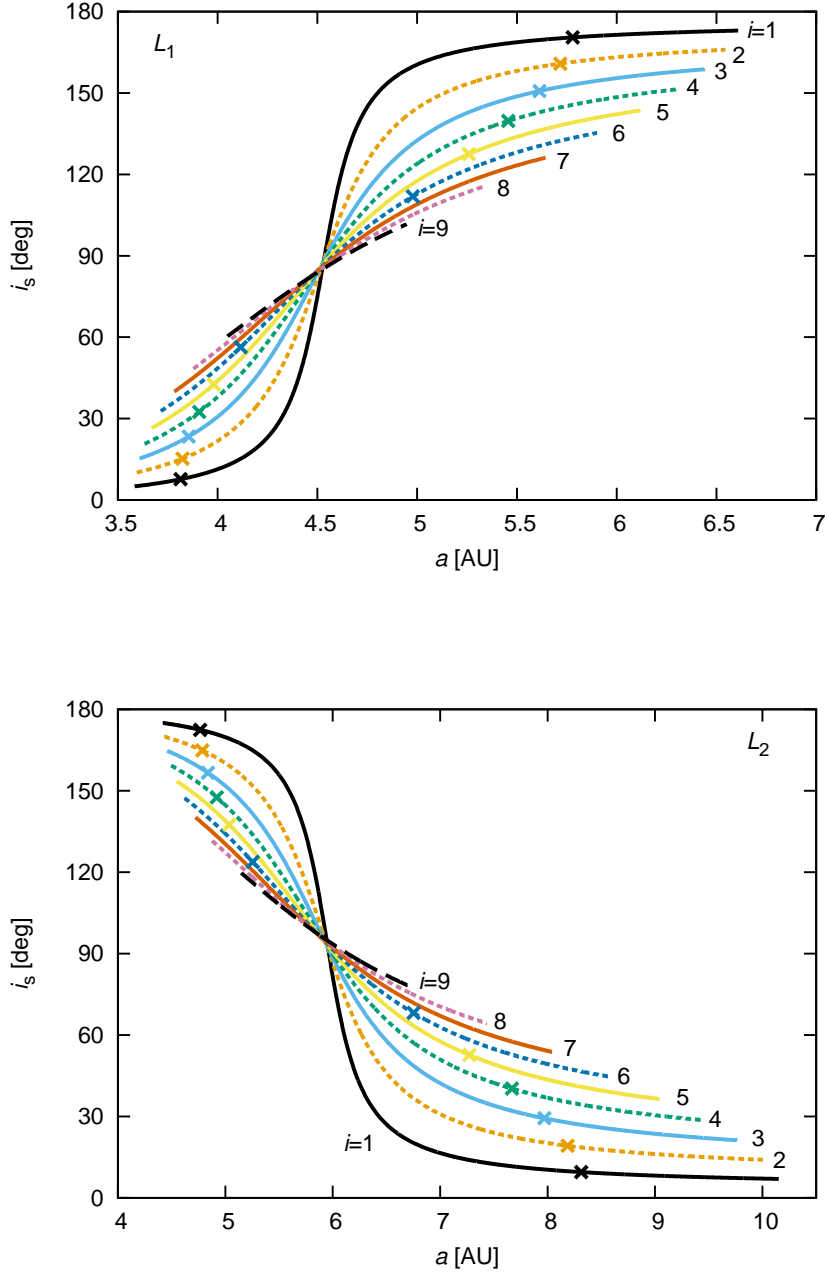}}
    \caption{
      Satellite inclination $i_{\rm s}$ of the planetocentric orbits as
      a function of heliocentric semimajor axis $a$ 
      for various heliocentric inclinations $i$, 
      given by Equation (\ref{eq:tc_id2}).
      The upper and lower panels are for $L_1$ and $L_2$ capture, respectively.
      The numbers labeled on the curves represent the values of $i$.
      Two crosses in each curve show the points of $\kappa=1$ capture.
      The solutions between the two crosses on each curve correspond to
      $\kappa<1$ capture.}
    \label{fig:tc_a_is}
  \end{center}
\end{figure}

The instantaneous inclination
 of the planetocentric orbits ("satellite inclination") is
 computed from the angular momentum.
The satellite's angular momentum at the $L_1$ or $L_2$ points 
 ${\bf r}_{\rm s}=(r_{\rm s}, 0, 0)$ with the velocity ${\bf v}_{\rm s}=(v_{{\rm s}, x},v_{{\rm s}, y},v_{{\rm s}, z})$
is written as 
\begin{eqnarray}
  {\bf h}_{\rm s}
  &=&{\bf r}_{\rm s}\times{\bf v}_{\rm s} 
  \nonumber\\
  &=&
  \left(
    \begin{array}{lll}
      0\\-r_{\rm s}v_{{\rm s},z}\\r_{\rm s}v_{{\rm s},y}
    \end{array}
  \right)
   \nonumber\\
  &=&
  h_{\rm s} \left(
    \begin{array}{lll}
      \sin\Omega_{\rm s}\sin i_{\rm s}\\
      -\cos\Omega_{\rm s}\sin i_{\rm s}\\
      \cos i_{\rm s}\\
    \end{array}
  \right),
\end{eqnarray}
 where the subscripts $x,y,z$ denote the $x,y,z$ components
 and $\Omega_{\rm s}$ is the longitude of the ascending node
 such that $\cos\Omega_{\rm s}=-1$ and 0 for $L_1$ and $L_2$ captures, respectively.  
Then,
\begin{eqnarray}
  \tan i_{\rm s}&=&
  \left\{
    \begin{array}{ll}
      \frac{-v_{{\rm s},z}}{v_{{\rm s}, y}} & {\rm for} \; L_1 {\rm \;capture},
      \\
      \frac{v_{{\rm s},z}}{v_{{\rm s}, y}} & {\rm for} \; L_2 {\rm \;capture}.
      \\
    \end{array}
  \right.
  \label{eq:tc_id}
\end{eqnarray}
Substituting 
 ${\bf v}_{\rm s}={\bf v}_{\rm rel}$ 
 into Equation (\ref{eq:tc_id}), we have
\begin{eqnarray}
  \tan i_{\rm s}&=&
  \left\{
    \begin{array}{ll}
      \frac{\chi\sin i }{1-\chi\cos i}  & {\rm for} \; L_1 {\rm \;capture},
      \\
      \frac{\chi\sin i}{\chi\cos i -1}  & {\rm for} \; L_2 {\rm \;capture}.
      \\
    \end{array}
  \right.
  \label{eq:tc_id2}
\end{eqnarray}
As already shown, the heliocentric $i$ of the captured bodies
is limited to $\lesssim 10^\circ$. 
However, note that Equation (\ref{eq:tc_id2}) shows that 
the planetocentric $i_{\rm s}$ can take any value. 
Figure \ref{fig:tc_a_is} shows $i_{\rm s}$ as a function of $a$ for 
various $i<i_{\rm max}$.
The region between the two crosses on each curve is for capture at apocenter
 (i.e., $\kappa=1-e_{\rm s}<1$).

The transition between prograde and retrograde captures occurs at
$i_{\rm s} = 90^\circ$. 
From Equation (\ref{eq:tc_id2}), the heliocentric semimajor axis for this transition, $a_{90}$, 
is derived by $\chi = 1/\cos i$, that is,
\begin{eqnarray}
  \frac{a_{90}}{a_{\rm p}} = \left(\frac{2}{A_\pm}-\frac{1}{\cos^2 i} \right)^{-1}.
  \label{eq:tc_xxxx}
\end{eqnarray}
In the limit that $i\rightarrow 0$, this condition is reduced
 to $a/a_{\rm p}\rightarrow B_\mp$ .
The values of $a_{90}$ for $i=0$ for the eight planets are presented in Table \ref{tb:tc_range} with 
 $a_{\rm min}$ and $a_{\rm max}$ for $i=0$.
As already shown, another planetocentric orbital element, $e_{\rm s}$,
 is determined by $a_{\rm s}(1 \pm e_{\rm s}) = r_{\rm H}$
 (Eq. (\ref{eq:tc_rs})).

\subsection{The Satellite's Inclination Distribution}
\label{ss:tc_sod}

Using Equations (\ref{eq:tc_hokaku}) and (\ref{eq:tc_id2}), 
we generate the planetocentric orbital distributions of satellites
for Jupiter, $m_{\rm p} = 9.55\times 10^{-4}M_\odot$ and $a_{\rm p}=5.2$AU,
 assuming that the heliocentric $a-i$ distribution is uniform in the ranges
 $a_{\rm min}<a<a_{\rm max}$ and $i < i_{\rm max}$.
We compute the distributions for $L_1$ and $L_2$ captures and add them.

The top panel in Figure \ref{fig:tc_bu_is_na} shows the $i_{\rm s}$ distribution for
 various ranges of the heliocentric semimajor axis $a$.
The vertical axis indicates the fraction of bodies in each $a$ bin
 with a width of 0.5 AU.
Basically, the behavior is similar to the relation in the case $i=0$
 summarized in Equation (\ref{eq:tc_12345}); 
 for the middle $a$ range near $a_{\rm p}$ (i.e., $4.2<a<6.2$ AU), 
 the distribution is dominated by retrograde orbits, 
 whereas the $a$ range on both sides of it (i.e., $a<4.2$ AU and $a>6.2$) is
 dominated by prograde orbits.
The peak of the $i_{\rm s}-$ distribution shifts outward as a function of $a$  
 for $a<5.2$ AU and inward for $a>5.2$ AU.
The lowest values of $i_{\rm s}$ are obtained for $a$ farthest from 
 $a=5.2$ AU; this can be explained by
 the small $i$ allowed for the capture as shown in
 Equations (\ref{eq:tc_a_mn}) and (\ref{eq:tc_a_mx}) and Figure \ref{fig:tc_a_i}.
With small $i$, 
 the orbital behavior around these regions is similar to that of the coplanar case
 i.e.,  $i_{\rm s}$ is close to 0 or 180$^\circ$.
The distribution of some $a$ regions that satisfy the temporary capture condition at both $L_1$ 
 and $L_2$ has a secondary peak.

Note that the top panel of Figure \ref{fig:tc_bu_is_na} is
 generated using the assumption that 
 $L_1$ and $L_2$ captures occur with the same probability
 and that $\kappa$ uniformly ranges from 0 to 2.
If we consider temporary capture as the origin of irregular satellites,
 orbits with $\kappa <1$ should be more favorable 
 because 
 orbits with $\kappa < 1$ correspond to tightly bound orbits
 and those with $\kappa > 1$ correspond to elongated satellite 
 orbits with their apocenter outside the Hill sphere of the planet.
The numerical calculations in the next section
clearly show this trend.

\section{Comparison with Numerical Results}

We perform numerical calculations for temporary capture of bodies
 by Jupiter to evaluate the relevance of our analytical formulae.
We will show that the dependence of prograde/retrograde capture
on the heliocentric semimajor axis of asteroids predicted by
our formulae agrees with the numerical results.

\subsection{Methods and Initial Conditions}

We compute the orbital evolution of $5\times 10^4$ bodies perturbed by Jupiter 
 moving along a circular orbit, using a 4th order Hermite integration scheme
 for $10^6$ years or less.
In our analytical derivation (Section 2), the three-body problem was
split into two problems of two bodies, Sun-asteroid and Jupiter-asteroid.
Here, because we use 
 numerical orbital integration,
we consider the circular restricted three-body problem (Sun-Jupiter-asteroid),
 just as previous investigators have (see the Introduction).

We consider asteroids to initially be uniformly distributed on the $a-T$ plane 
 between $a_{\rm min}<a<a_{\rm max}$ and $T_{\rm min}<T<T_{\rm max}$.
We randomly choose $i<i_{\rm max}$.
The minimum and maximum values of $a$, $T$, and $i$ and Jupiter's semimajor axis and mass
 used in the calculation are given in Table \ref{tb:tc_range}.

We count asteroids as temporary captures
 if they satisfy two conditions:
 (1) they must stay within 3 $r_{\rm H}$ from Jupiter longer than one orbital period of Jupiter
 and (2) the minimum distance from Jupiter be less than 1 $r_{\rm H}$.
If an asteroid collides with Jupiter or the Sun, or has $e>1$ at $r>30$ AU,
 it is removed from the calculation.
We output the heliocentric orbital elements of the temporarily captured asteroids 
 at 3 $r_{\rm H}$ away from Jupiter before and then after the temporary capture.
We also output the planetocentric orbital elements 
 when 1 $r_{\rm H}$ away from Jupiter before and after the temporary capture.
All the planetocentric orbital elements are calculated within a two-body framework that consists of
 Jupiter and the asteroids.

\subsection{Results}

\subsubsection{Incident parameters to the Hill sphere}
\begin{figure}[hbtp]
  \begin{center}
    \resizebox{13cm}{!}{\includegraphics{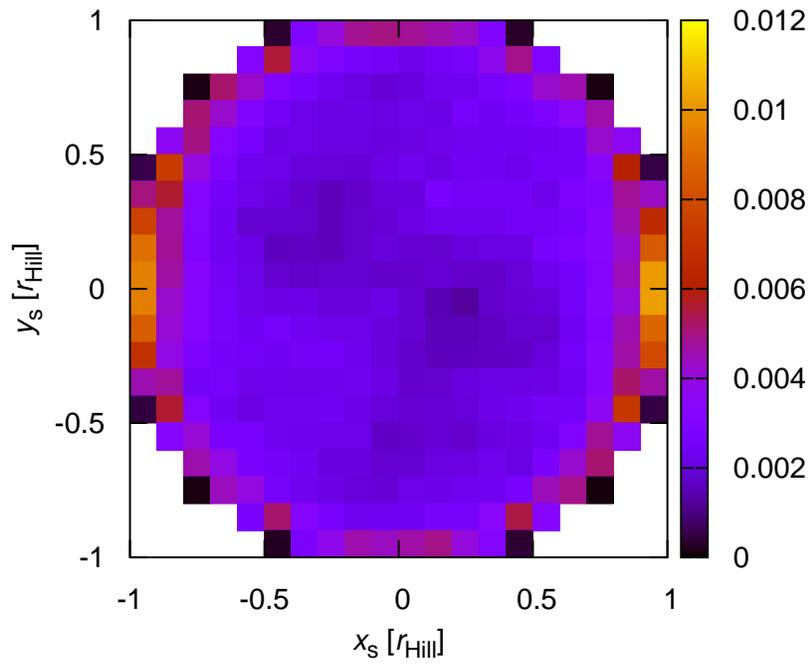}}
    \caption{Location of the bodies on the Hill sphere of Jupiter on the $x_s-y_s$ plane, at the moment 
      when the bodies enter the Hill sphere for the first time during each temporary capture.}
    \label{fig:tc_xy}
  \end{center}
\end{figure}

\begin{figure}[hbtp]
  \begin{center}
    \resizebox{13cm}{!}{\includegraphics{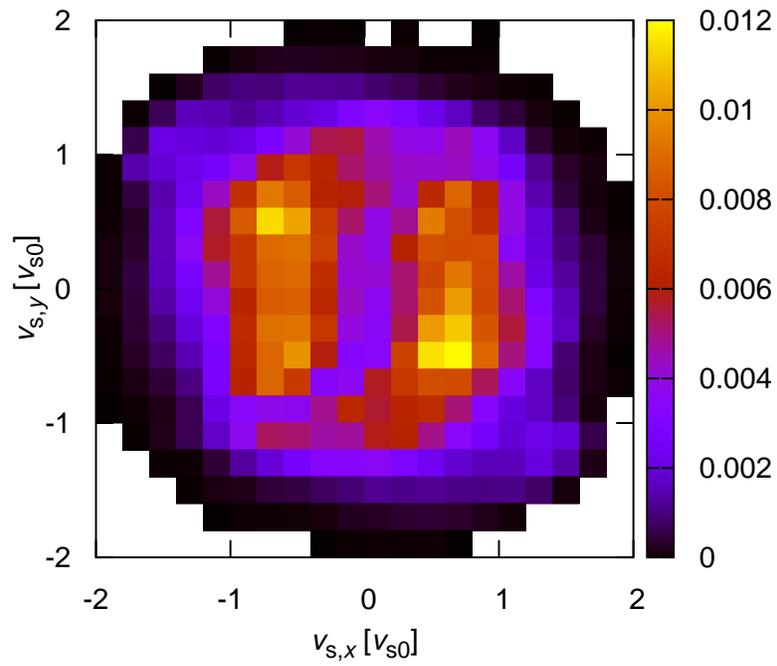}}
    \caption{Velocity distribution of the bodies on the Hill sphere of Jupiter on the 
      $v_{{\rm s}, x}-v_{{\rm s}, y}$
      plane, at the moment 
      when the bodies enter the Hill sphere for the first time during each temporary capture.
      The values are scaled by the circular veocity of the satellite at $1r_{\rm H}$ away from Jupiter.
    }
    \label{fig:tc_vv}
  \end{center}
\end{figure}

\begin{figure}[hbtp]
  \begin{center}
    \resizebox{13cm}{!}{\includegraphics{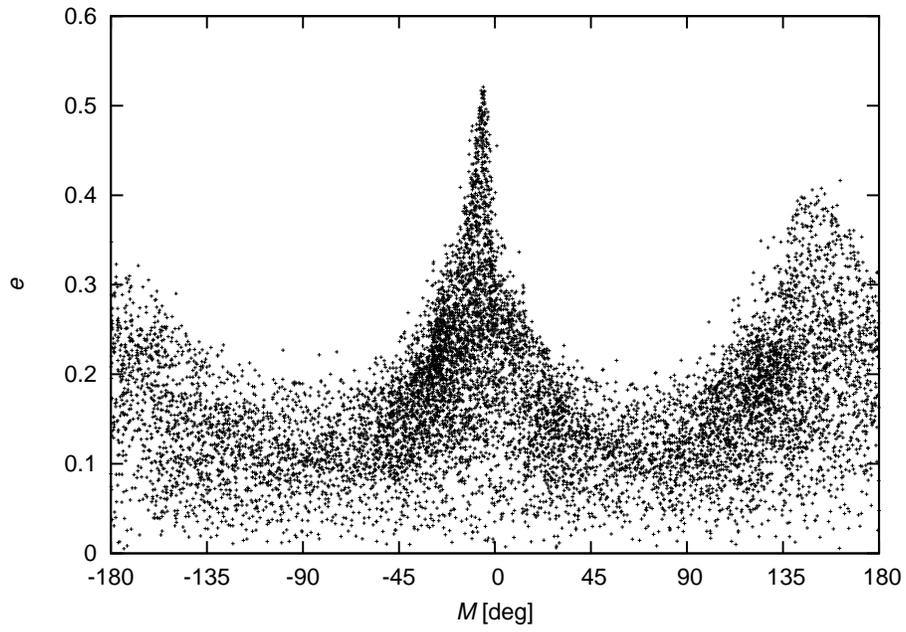}}
    \caption{Distribution of the mean anomaly $M$ and eccentricity of the bodies 
      in heliocentric orbit, at the moment 
      when the bodies enter the Hill sphere for the first time during each temporary capture
      (the same as Fig. \ref{fig:tc_xy}).}
    \label{fig:tc_le}
  \end{center}
\end{figure}

We have found 
$1.6\times 10^4$ temporary captures by Jupiter during the calculation.
Figure \ref{fig:tc_xy} shows the two-dimensional (2D) distribution of 
 the positions of the captured bodies as they 
 enter Jupiter's Hill sphere for the first time during each temporary capture.
The two concentrations at $(x_{\rm s}, y_{\rm s})=(-1, 0)$ and $(1, 0)$ 
 correspond to $L_1$ and $L_2$, showing that condition [2] is approximately valid.
The concentration around the perimeter is a geometrical effect.
Condition [1] corresponds to $|v_{{\rm s},x}|,|v_{{\rm s}, y}|<v_{\rm sat}$.
Figure \ref{fig:tc_vv} show the 2D distribution of the incident velocities of the 
 captured bodies
 on the $v_{{\rm s}, x}-v_{{\rm s}, y}$ plane at the same moment as Figure \ref{fig:tc_xy}.
The values of $v_{{\rm s}, x}$ and $v_{{\rm s}, y}$ are scaled by 
 by the circular velocity of the satellite at $1r_{\rm H}$ away from Jupiter,
 $v_{\rm s0}=\sqrt{Gm_{\rm J}/r_{\rm H}}$, where $m_{\rm J}$ is Jupiter's mass. 
Since $|v_{{\rm s},x}|,|v_{{\rm s}, y}|<v_{\rm s0}$ is mostly satisfied,
 condition [1] is approximately valid.
The upper-left and lower-right peaks in Figure \ref{fig:tc_vv} correspond to 
 the two concentrations at $(x_{\rm s}, y_{\rm s})=(1, 0)$ and $(-1, 0)$ in Figure \ref{fig:tc_xy},
 which are at $L_2$ and $L_1$, respectively.
Condition [3] corresponds to $|v_{{\rm s},x}|\gg|v_{{\rm s}, y}|$, 
while the numerical orbital integrations
 show that $|v_{{\rm s}, x}|\sim|v_{{\rm s}, y}|$.
The three-body effect, which we do not take into account,
 is important near $r_{\rm s}=r_{\rm H}$
 and planet's gravitational pull to heliocentric orbits causes a non-zero value
 of $|v_{{\rm s}, x}|$.
Since $|v_{{\rm s}, x}|$ is not larger than $|v_{{\rm s}, y}|$, condition [4] would not be invalid.
Figure \ref{fig:tc_le}
 shows the distribution of the mean anomaly $M$ and eccentricity
 in the heliocentric distance at the same moment for Figure \ref{fig:tc_xy}.
The two concentrations around $M\lesssim 0$ and $M\lesssim 180^\circ$ correspond to 
 the aphelion and perihelion, showing that condition [3] is approximately valid.
These concentrations are not found for small $e$, since the radial velocity is
 small even in the orbital phases far from apocenter and pericenter.

\subsubsection{Distributions of planetocentric inclinations}

\begin{figure}[hbtp]
  \begin{center}
    \resizebox{13cm}{!}{\includegraphics{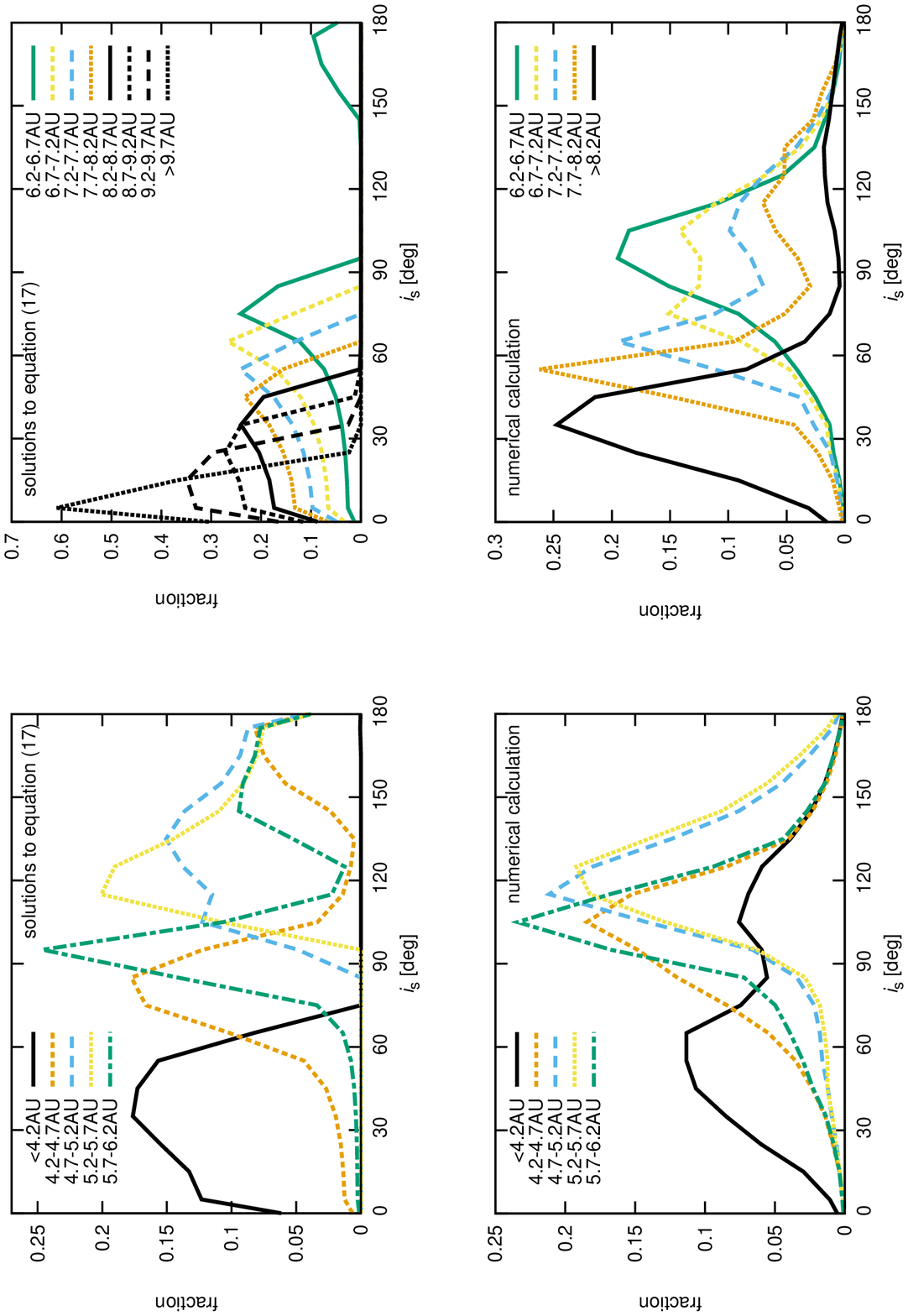}}
    \caption{Distribution of satellite orbital inclination $i_{\rm s}$ obtained from Eqs. (\ref{eq:tc_hokaku}) 
      and (\ref{eq:tc_id2}) assuming a uniform distribution on the $a-i$ plane for various $a$ (upper panels),
      compared with results from numerical calculation (lower panels).}
    \label{fig:tc_bu_is_na}
  \end{center}
\end{figure}

The lower panel in Figure \ref{fig:tc_bu_is_na} shows the 
 $i_{\rm s}$ distribution of the captured bodies
 when they cross
 the Jovian Hill sphere for the first time during each temporary capture
 (the same timing as Figure \ref{fig:tc_xy}), as a function of
 the heliocentric semimajor axes just before (3 $r_{\rm H}$ away from Jupiter) 
 the temporary  capture ($a_{\rm tc}$).
The distribution is scaled for individual $a_{\rm tc}$ bins.
The relative frequency distribution among different $a_{\rm tc}$ is described in the next section.

 The peaks of the distribution are shifted depending on $a_{\rm tc}$.
The $a_{\rm tc}$ dependence is similar to 
 the distribution generated in Section \ref{ss:tc_sod} using Equations (\ref{eq:tc_hokaku}),
 (\ref{eq:tc_id}), and (\ref{eq:tc_id2}).
The results of the numerical calculations share the following common features with
 the analytical predictions 
(the upper panel of Fig.~\ref{fig:tc_bu_is_na}),
(1) Bodies originating from heliocentric semimajor axes at 4.7 AU $<a_{\rm tc}<$ 5.7 AU around Jupiter's orbit
generally produce retrograde satellite orbits with the highest values of $i_{\rm s}$
when they are captured,  and 
(2) Prograde orbiters with small $i_{\rm s}<30^\circ$ mostly 
  come from the regions relatively far from Jupiter's orbit, $a_{\rm tc}<4.2$ AU or $a_{\rm tc}>8.2$ AU.
These features are also present in the results of the planar case in section 2.2,
so that the basic dynamics for these are explained in section 2.2.

On the other hand, 
 the analytical distribution (the upper panel of Figure \ref{fig:tc_bu_is_na})
 differs from our numerical results in the following ways:
 (1) The secondary peaks of 4.2 AU$<a_{\rm tc}<$6.7 AU at $i_{\rm s}\gtrsim 150^\circ$
 in the analytical distribution are not found in the numerical distribution, 
 while (2) the distributions for $a_{\rm tc}<$ 4.2 AU and $a_{\rm tc}>$ 6.7 AU are
 bimodal in the numerical distribution, but this was not predicted analytically.
The peaks in (1) correspond to bodies captured at pericenter.
As we anticipated in section 2.3, temporary captures at the pericenters of satellite orbits are infrequent,
although such captures do exist.
This feature is enhanced when we investigate longer 
 temporary captures (e.g., $>$ 100 years).
The bimodal distribution in (2) is probably due to
 differences between $a$ in Equation (\ref{eq:tc_hokaku}) and the numerically obtained $a_{\rm tc}$,
caused by the assumption of splitting the restricted three-body problem
into a pair of independent two-body problems that we have adopted to derive Equation (\ref{eq:tc_hokaku}).

\subsubsection{Capture frequency as a function of the heliocentric semimajor axes of asteroids}

The relative frequency distribution among different $a_{\rm tc}$ obtained 
by the numerical orbital integration is shown by
 the black histograms in Figure \ref{fig:tc_bu_appa_na}.
In section 2.2, we predicted that temporary capture occurs only for asteroids
 between  3.6 AU $< a <10.2$ AU. 
The numerical result is consistent with this prediction.
However, the distribution is strongly skewed toward smaller $a$ within 
the region capable of capture, although
we spread the initial heliocentric semimajor axes of asteroids uniformly.

The peak at relatively small $a$ (4.2 AU $<a_{\rm tc}<$ 4.7 AU) may be 
 due to the short orbital periods of the inner orbits
 and some stable regions in mean-motion resonances with Jupiter 
 such as the 3:2 Hilda asteroids.
This region corresponds to 
 the $a_{\rm tc}$ bin producing the peak at $i_{\rm s}\simeq 110^\circ$.
This may explain why giant planets have retrograde irregular satellites more often 
than prograde ones.
The frequency for $a_{\rm tc}<4.2$ AU is much larger than that for $a_{\rm tc}>7.7$ AU.
This means that the prograde orbiters with small $i_{\rm s}<30^\circ$ are 
 mainly from the inner region. 

The orange and blue histograms in Figure \ref{fig:tc_bu_appa_na} show the results 
 of additional numerical calculations with an eccentric Jupiter 
($e=0.05$) instead of 
 a circular Jupiter and with an extra perturbation from Saturn ($a_{\rm p}=9.55$ AU, 
 $m_{\rm p}=2.86\times 10^{-4}M_\odot$; cf. \citet{kd96}). 
Neither the Jupiter's eccentricity nor Saturn's presence changes the overall features
 of the $a_{\rm tc}$ frequency distribution.

\section{Summary and Discussion}

To discuss the temporary capture of asteroids by a planet,
we have investigated the dependence of prograde/retrograde 
(inclinations of the resultant satellite orbits) capture of asteroids on
their original heliocentric semimajor axes
using analytical arguments and numerical orbital integrations.
In the orbital integrations, we solved the circular restricted
three-body problem (Sun-Jupiter-asteroid).  
In the analytical arguments, we split the three-body problem
into two independent systems of two-body problems (Sun-asteroid
and planet-satellite), where the planetary semimajor axis and mass are
scaled and the arguments are not specific to Jupiter.
The two systems are combined by identifying the relative
velocity between Jupiter's heliocentric circular orbit
 and the asteroid's heliocentric Keplerian eccentric orbit with
 a planetocentric Keplerian eccentric orbit as a satellite at 
 the $L_1$ or $L_2$ points of the planet's Hill sphere. 

We have found a clear dependence of prograde/retrograde capture
on the pre-capture heliocentric semimajor axes of the asteroids.
Capture is mostly retrograde for the asteroids 
from orbits near the planetary orbit,
more specifically, from heliocentric semimajor axes $a$ in the range
\begin{equation}
\frac{1-\hat{r}_{\rm H}}{1+\hat{r}_{\rm H}} \la \frac{a}{a_{\rm p}} \la 
\frac{1+\hat{r}_{\rm H}}{1- \hat{r}_{\rm H}},
\label{eq:retro}
\end{equation}
where $a_{\rm p}$ is the planet's semimajor axis, $\hat{r}_{\rm H}=(m_{\rm p}/3M_\odot)^{1/3}$,
and $m_{\rm p}$ is the planetary mass.
On the other hand, capture is mostly prograde for those asteroids 
from orbits far from the planetary orbit,
\begin{equation}
\frac{a}{a_{\rm p}} \la
\left\{
    \frac{2}{1+\hat{r}_{\rm H}}
    -\left[1 -
      \sqrt{6}\hat{r}_{\rm H}\right]^2
    \right\}^{-1}
    \mbox{ or }
\frac{a}{a_{\rm p}} \ga
\left\{
    \frac{2}{1-\hat{r}_{\rm H}}
    -\left[1 +
      \sqrt{6}\hat{r}_{\rm H}\right]^2
    \right\}^{-1}.  
\label{eq:prog}
\end{equation}
We also found that asteroids at 
    $a/a_{\rm p} \la \left\{\frac{2}{1-\hat{r}_{\rm H}}
    -\left[1 -  \sqrt{6}\hat{r}_{\rm H}\right]^2\right\}^{-1}$
    and $a/a_{\rm p} \ga \left\{\frac{2}{1+\hat{r}_{\rm H}}
    -\left[1 + \sqrt{6}\hat{r}_{\rm H}\right]^2\right\}^{-1}$
or those with heliocentric orbital inclinations larger than 10 degrees 
cannot be captured.
The conditions (\ref{eq:retro}) and (\ref{eq:prog}) are
 come from the analytical arguments.
The numerical orbital integrations show similar results, 
although the prograde/retrograde boundaries are less clear.

Our results indicate that retrograde irregular satellites
  are most likely to be captured bodies from the orbits near the host planet's 
 orbit, 
  whereas most prograde irregular satellites originate from farther regions
  on either side of the host planet.
Note that our numerical results show that the capture probability 
  is much higher for bodies from inner regions than for outer ones.
Therefore, the prograde region is actually more concentrated than the
retrograde region.

These results suggest that, in Jupiter's case, 
  the retrograde irregular satellites likely originated as
  Trojan asteroids and
  the majority of the prograde irregular satellites are
  from far inner regions such as main-belt asteroids.
This is consistent with the recent observations of 
  irregular satellites and Trojan asteroids of Jupiter.
\citet{snckh00} found differences between 
    prograde and retrograde groups
    from near-infrared observations of six of the 
    eight known Jovian irregular satellites detected
    in the Two-micron All Sky Survey.
    They suggested that the retrograde satellites exhibit
    much greater diversity among themselves than the
    prograde satellites 
    and that the retrograde (prograde) satellites 
    may be similar to D-type (C-type) asteroids, 
    although their samples are only of eight objects.
BVR photometry of Jovian irregular satellites
    presented by \citet{rwc01} and \citet{ghga03} shows
    a concentration of prograde satellites in 
    a small region, except Themisto,
    and a redder and more diverse distribution for
    retrograde satellites 
    on the $B-V$ versus $V-R$ color-color plot. 
    The region of the diversity of retrograde satellites
    matches that of Trojan asteroids given in \citet{hbp12}.

\begin{figure}[hbtp]
  \begin{center}
    \resizebox{13cm}{!}{\includegraphics{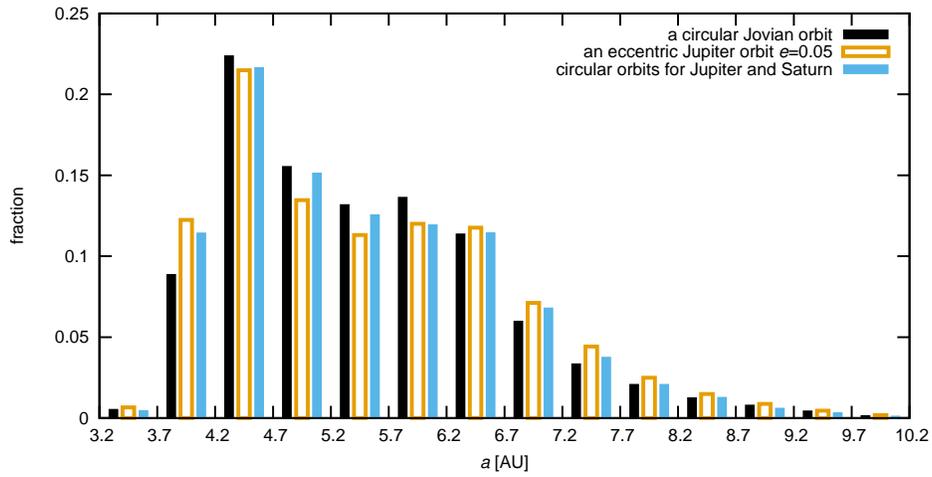}}
    \caption{Distribution of $a_{\rm tc}$ obtained by numerical calculations. 
      Black: with a circular Jovian orbit;
      open orange: with an eccentric Jupiter orbit ($e=0.05$, additional calculations); 
      blue: with circular orbits for Jupiter and Saturn (additional calculations).}
    \label{fig:tc_bu_appa_na}
  \end{center}
\end{figure}

Small eccentricity ($e=0.05$) for Jupiter made little 
 difference in the results with a circular Jupiter case 
 (Fig. \ref{fig:tc_bu_appa_na}).
 However, the effect of eccentricity, especially for less massive planets 
 such as Mars, is not negligible.
 In our next paper we will expand the results to eccentric planet cases.

\acknowledgements
We thank Takayuki Tanigawa for valuable discussions.
We also thank an anonymous referee for his/her useful comments that helped to improve the paper.
This work was supported by JSPS KAKENHI grant Number 23740335.
Data analyses were in part carried out on the PC cluster at 
 the Center for Computational Astrophysics, 
 National Astronomical Observatory of Japan.

\begin{table}[hbtp]
  \begin{center}
    \scalebox{0.8}{
      \begin{tabular}{ccc|ccccccc}
	\hline\hline
	Planet&$a_{\rm p}$ (AU)&$m_{\rm p}$ ($M_\odot$)& $i_{\rm max}$ (degree)&
	$L_1/L_2$ & $a_{\rm min}$ (AU)& $a_{90}$ (AU)& $a_{\rm max}$ (AU)& $T_{\rm min}$& $T_{\rm max}$\\
	\hline
        Mercury  & 0.387  & 1.66e-07 & 0.5348  & $L_1$ & 0.3771 & 0.384062& 0.3913 & 2.99987 & 3.00004 \\ \cline{6-10}
         & & &  & $L_2$ & 0.3828 & 0.389961& 0.3975 & 2.99987 & 3.00004 \\ \hline
        Venus  & 0.723  & 2.45e-06 & 1.312  & $L_1$ & 0.6794 & 0.709609 & 0.7434 & 2.99922 & 3.00026 \\ \cline{6-10}
         & & &  & $L_2$ & 0.7042 & 0.736644 & 0.7731 & 2.99922 & 3.00026 \\ \hline
        Earth  & 1  & 3.00e-06 & 1.404  & $L_1$ & 0.9358 &0.980198& 1.03 & 2.99911 & 3.0003 \\ \cline{6-10}
         & & &  & $L_2$ & 0.9722 & 1.0202& 1.075 & 2.99911 & 3.0003 \\ \hline
        Mars  & 1.52  & 3.72e-07 & 0.6999  & $L_1$ & 1.47 & 1.50492& 1.542 & 2.99978 & 3.00007 \\ \cline{6-10}
         & & &  & $L_2$ & 1.498 & 1.53524& 1.574 & 2.99978 & 3.00007 \\ \hline
        Jupiter  & 5.2  & 9.55e-04 & 9.628  & $L_1$ & 3.579 & 4.53527&6.632 & 2.95919 & 3.01467 \\ \cline{6-10}
         & & &  & $L_2$ & 4.412 & 5.96215 & 10.2 & 2.95792 & 3.01339 \\ \hline
        Saturn  & 9.55  & 2.86e-04 & 6.425  & $L_1$ & 7.307 & 8.71559&11.11 & 2.98163 & 3.00646 \\ \cline{6-10}
         & & &  & $L_2$ & 8.497 & 10.4643& 14.12 & 2.98125 & 3.00608 \\ \hline
        Uranus  & 19.2  & 4.37e-05 & 3.43  & $L_1$ & 16.46 & 18.2845& 20.72 & 2.99472 & 3.00182 \\ \cline{6-10}
         & & &  & $L_2$ & 17.97 & 20.1613 & 23.16 & 2.99466 & 3.00176 \\ \hline
        Neptune  & 30.1  & 5.15e-05 & 3.623  & $L_1$ & 25.61 & 28.5861 & 32.63 & 2.99411 & 3.00203 \\ \cline{6-10}
         & & &  & $L_2$ & 28.08 & 31.6941 &36.74 & 2.99404 & 3.00196 \\ \hline
      \end{tabular}
    }
    \caption{Ranges of $i$, $a$, and $T$ from Equation (\ref{eq:tc_hokaku}) and $a_{90}$ from 
      Equation (\ref{eq:tc_xxxx}) for eight planets.}
    \label{tb:tc_range}
  \end{center}
\end{table}

\end{document}